\documentclass[a4paper]{jpconf}
\usepackage{graphicx}
\usepackage{amssymb}

\begin{document}

\title{The influence of magnetic field geometry on magnetars X-ray spectra}

\author{D Vigan\`o$^1$, N Parkins$^{2,3}$, S Zane$^3$, R Turolla$^{4,3}$, J A Pons$^1$ and J A Miralles$^1$}

\address{$^1$ Departament de Fisica Aplicada, Universitat d'Alacant, Ctra. San Vicent del Raspeig, s/n. 03690 San Vicent del Raspeig, Spain}
\address{$^2$ University of Liverpool, Liverpool L69 3BX, UK}
\address{$^3$ Mullard Space Science Laboratory, University College London, Holmbury St. Mary, Dorking, Surrey RH5 6NT, UK}
\address{$^4$ Universit\`a di Padova, Dipartimento di Fisica, via F. Marzolo 8, I-35131 Padova, Italy}

\ead{daniele.vigano@ua.es}

\begin{abstract}
Nowadays, the analysis of the X-ray spectra of magnetically powered neutron stars or {\it magnetars} is one of the most valuable tools to gain insight into the physical processes
occurring in their interiors and magnetospheres. In particular, the magnetospheric plasma leaves a strong imprint on the observed X-ray spectrum by means of Compton up-scattering of the thermal radiation coming from the star surface. Motivated by the increased quality of the observational data, much theoretical work has been devoted to develop Monte Carlo (MC) codes that incorporate the effects of resonant Compton scattering (RCS) in the modeling of radiative transfer of photons through the magnetosphere. The two key ingredients in this simulations are the kinetic plasma properties and the magnetic field (MF) configuration. The MF geometry is expected to be complex, but up to now only mathematically simple solutions (self-similar solutions) have been employed. In this work, we discuss the effects of new, more realistic, MF geometries on synthetic spectra. We use new force-free solutions \cite{vigano11} in a previously developed MC code \cite{ntz08b} to assess the influence of MF geometry on the emerging spectra. Our main result is that the shape of the final spectrum is mostly sensitive to uncertain parameters of the magnetospheric plasma, but the MF geometry plays an important role on the angle-dependence of the spectra.
\end{abstract}

\section{Introduction}
Neutron stars (NSs) can be powered by different sources of energy:
rotation, accretion, residual heat and magnetic field. Up to now,
more than twenty X-ray pulsating sources are identified as
isolated, slowly rotating ($P\sim 2-10~$s) neutron stars, whose
large X-ray luminosity  $L_x\sim 10^{33}-10^{35}~$erg/s cannot
usually be explained in terms of rotational energy, at variance
with rotation-powered pulsars (RPPs). Historically, they are
classified as Anomalous X-ray Pulsars (AXPs) or Soft Gamma
Repeaters (SGRs), and they are recognized as magnetars. In these objects, MF decay and evolution is the dominant energy source and crustal deformations produced by the huge internal magnetic stresses are the likely cause of bursts and episodic giant flares. In most cases, the spin-down inferred external magnetic field turns out to be super-strong. Recently, increasing evidence gathered in favor of a unified scenario for isolated neutron stars, their different observed manifestations being mainly due to a different topology and strength of the internal MF, and to evolution (\cite{pons11}).

The typical soft X-ray ($< 10~$keV) spectrum of magnetars is given by a thermal component ($kT\sim 0.3-0.5~$keV), plus a hard tail phenomenologically described by a power law with a photon index $\Gamma\sim 3-4$ \cite{mereghetti08}. This hard tail is believed to be produced by resonant compton (up)scattering (RCS) of photons emitted by the star surface onto mildly relativistic particles flowing in magnetosphere. In order to take into account for this effect, several codes have been developed in recent years \cite{fernandez07,ntz08a,ntz08b,fernandez11}. The two basic inputs are the external MF and a prescription for the particle space and velocity distribution. With these ingredients, and with a given energy distribution for the incoming surface photons (the seed spectrum), the MC simulations of RCS provide the emerging, reprocessed spectrum, which must be compared with observational data. 

Magnetars also exhibit a high-energy ($\sim 20$--200 keV) power-law tail, the origin of which is still under debate. One of the possible mechanisms is again RCS, whose efficiency at high energy is very sensitive to MF and plasma velocity. In this paper we discuss the effect of the geometry of the MF, which is obtained from new calculations of magnetostatic, force-free equilibria.

\section{Force-free twisted magnetosphere}\label{sec_forcefree}

The shape of the internal MF of NSs is largely unknown, but it is thought to be complex, with toroidal and poloidal components of similar strength, and with significant deviations from a dipolar geometry. In the crust, a large toroidal field is a necessary ingredient to explain the observed thermal and bursting properties of magnetars, even in the case in which they show a low value of the inferred external dipolar field \cite{turolla11}. The complex internal topology must be smoothly matched to an external solution, so that the internal MF geometry at the star surface has a strong effect on the external MF. In this framework, the RPP phenomenolgy is expected to be better explained by the presence of a NS with a simple, nearly potential, dipolar field, while magnetars activity is compatible with the presence of a complex external field, as a result of the transfer of magnetic helicity from the internal to the external field. A non-potential MF needs supporting currents, and charges can be either extracted from the surface or produced by one-photon pair creation in the strong MF \cite{beloborodov07}.

In the inner magnetospheric region MF lines corotate with the star, with footprints connected with the crustal field, which evolves on long timescales ($10^3-10^5$ years). Furthermore, resistive processes in the magnetosphere act on a typical timescale of years \cite{beloborodov09}, much longer than the typical response of the tenuous plasma, whose Alfv\'en velocity is near to the velocity of light. For these reasons, a reasonable approximation is to build stationary, force-free magnetospheric solutions, assuming that the footprints are anchored in the crust and neglecting resistivity. Then, the ideal MHD force-free equation is a simple balance between electromagnetic forces, provided that inertial, pressure, and gravity terms are negligible
\begin{equation}\label{eq_electromagnetic_forces}
 \rho_e \vec{E}+\frac{1}{c}\vec{J}\times\vec{B}=0~,
\end{equation}
where we have introduced the rotationally induced electric field $\vec{E}=-\vec{v}_{rot}\times \vec{B}$,  the related charge density $\rho_e$ and the current density $\vec{J}=c\vec{\nabla}\times\vec{B}/4\pi$.

The electric field gives corrections of order $(\rho \Omega /c)^2$. Neglecting such contributions is safe if we consider the region closest to a magnetar surface ($r\lesssim 100 r_\star$), and remembering that the light cylinder for slow rotators lies typically at a few thousands stellar radii ($r_l=c/\Omega$ and magnetar periods are in the 1--10~s range). Therefore, condition (\ref{eq_electromagnetic_forces}) reduces simply to $\vec{J}\parallel \vec{B}$: the particles can flow only along MF lines. Even if the pulsar equation simplifies in this limit, it still contains an arbitrary function, which basically describes the toroidal field as a function of the poloidal one (see \cite{vigano11} for the mathematical formulation). In the literature, only the simplest (semi-)analytical solutions have been derived and later employed in applications. In
particular, in the context of magnetars, the self-similar twisted dipole family of solutions is the most popular choice. Proposed for solar corona studies \cite{low90}, and later extended to the magnetar scenario \cite{tlk02}, in self-similar models the radial dependence of the MF components is described by a simple power law, $B_i\propto r^{-(p+2)}$. These solutions are semi-analytical and described by only one parameter, the radial index $p$ (see \cite{pavan09} for globally twisted multipolar fields). However, globally twisted solutions necessary evolve in time and magnetic configurations in which the twist affects only a limited bundle of field lines are likely required by observations \cite{beloborodov09}.

\begin{figure}
\centering
\includegraphics[width=.23\textwidth]{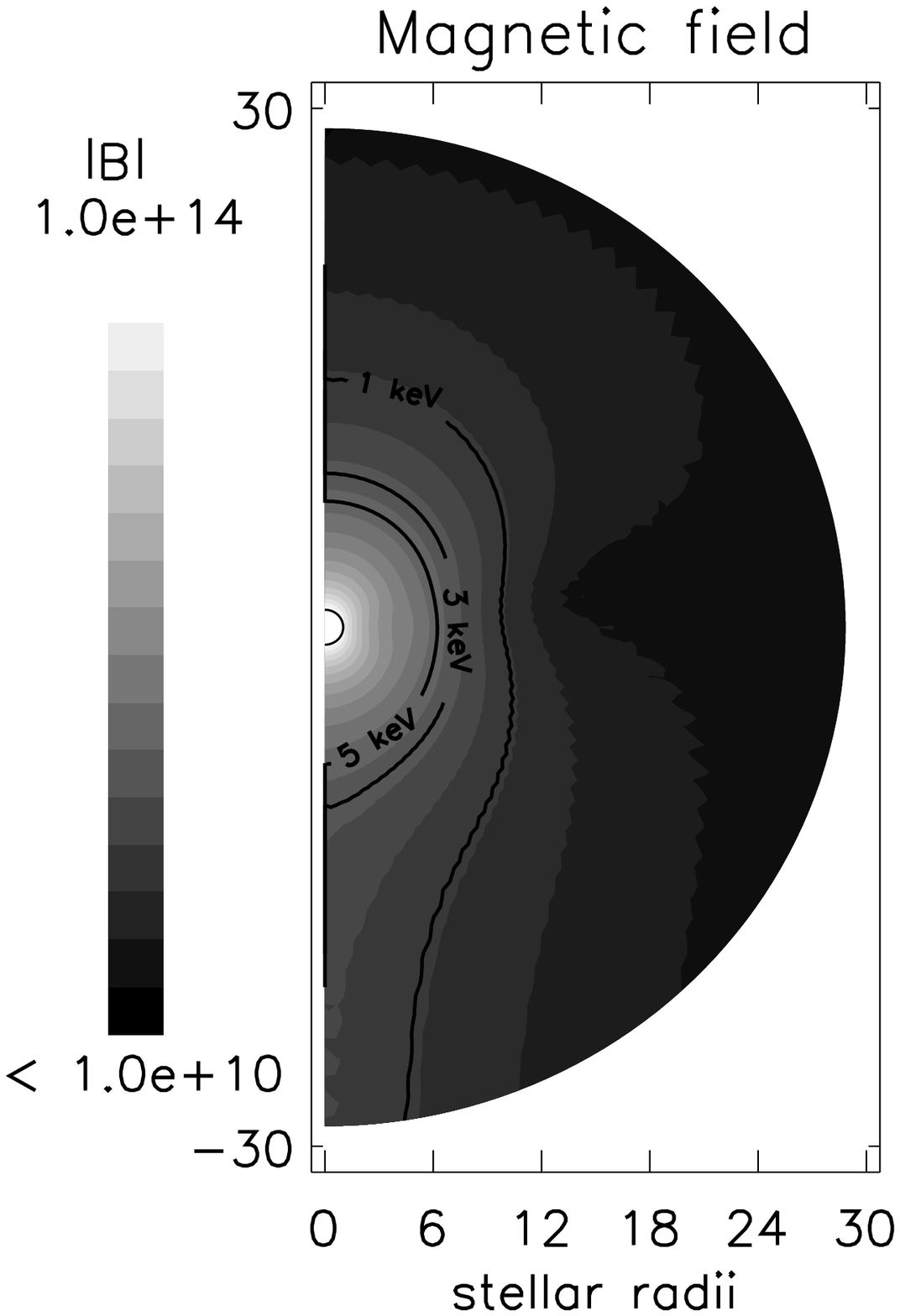}
\includegraphics[width=.23\textwidth]{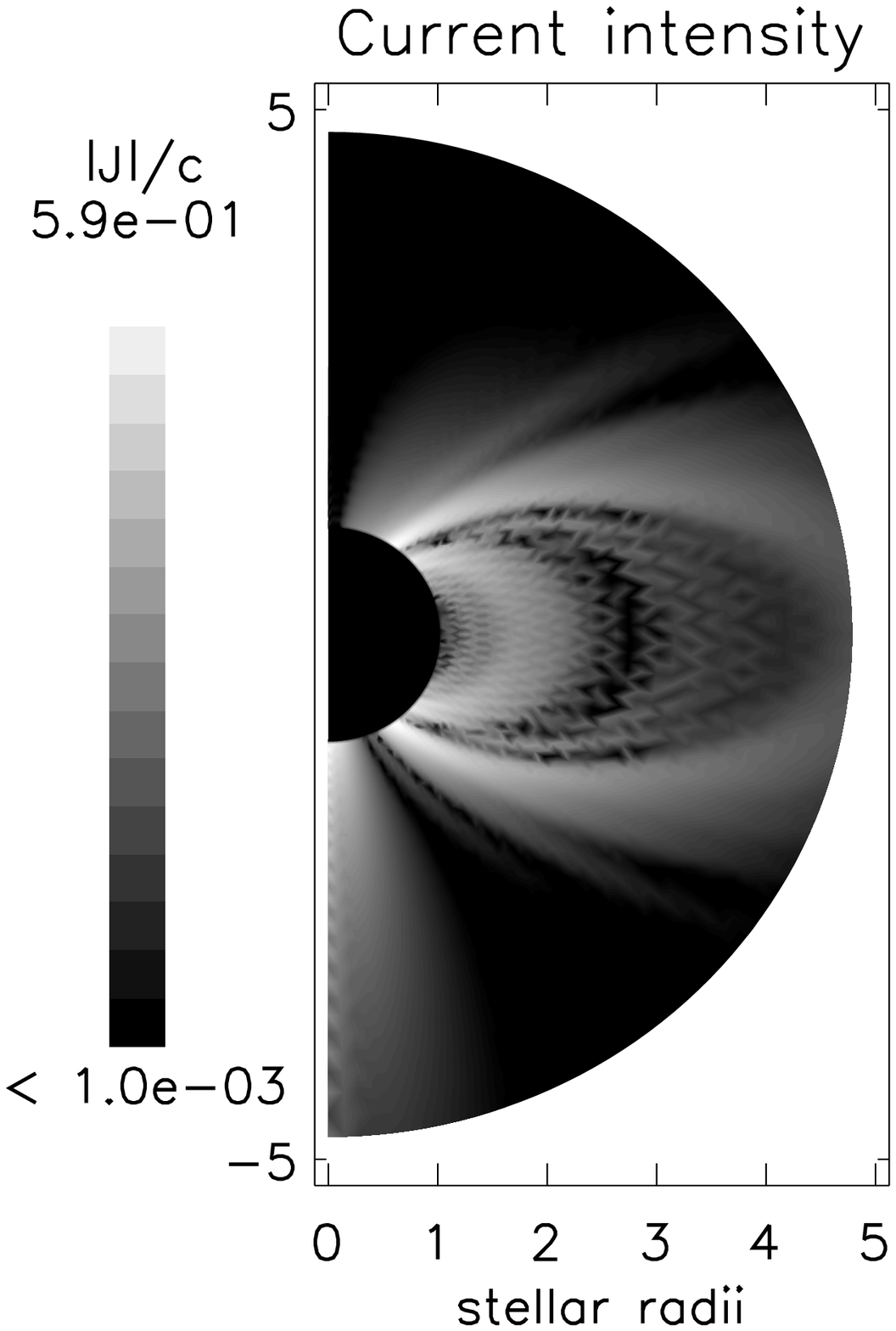}
\includegraphics[width=.23\textwidth]{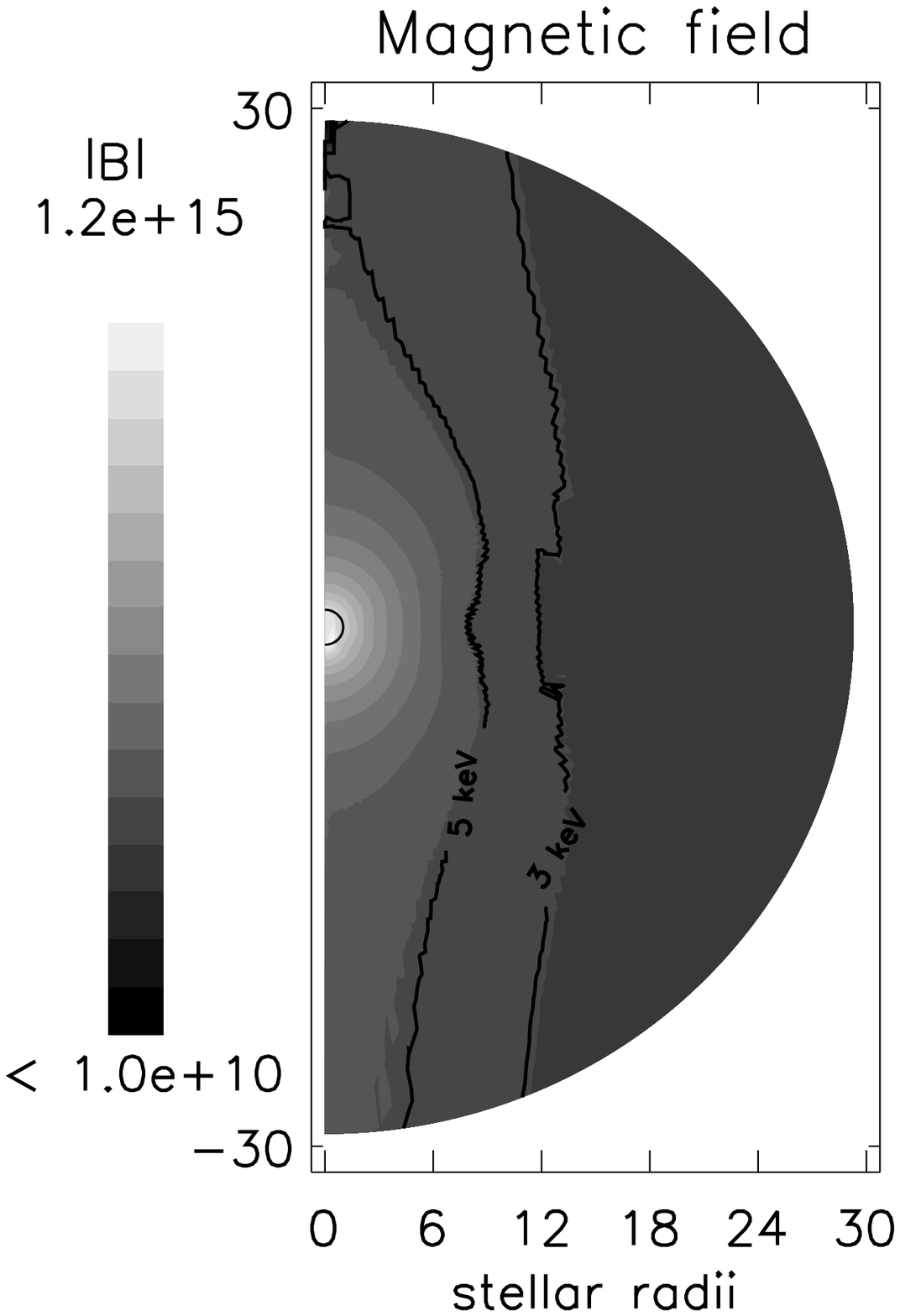}
\includegraphics[width=.23\textwidth]{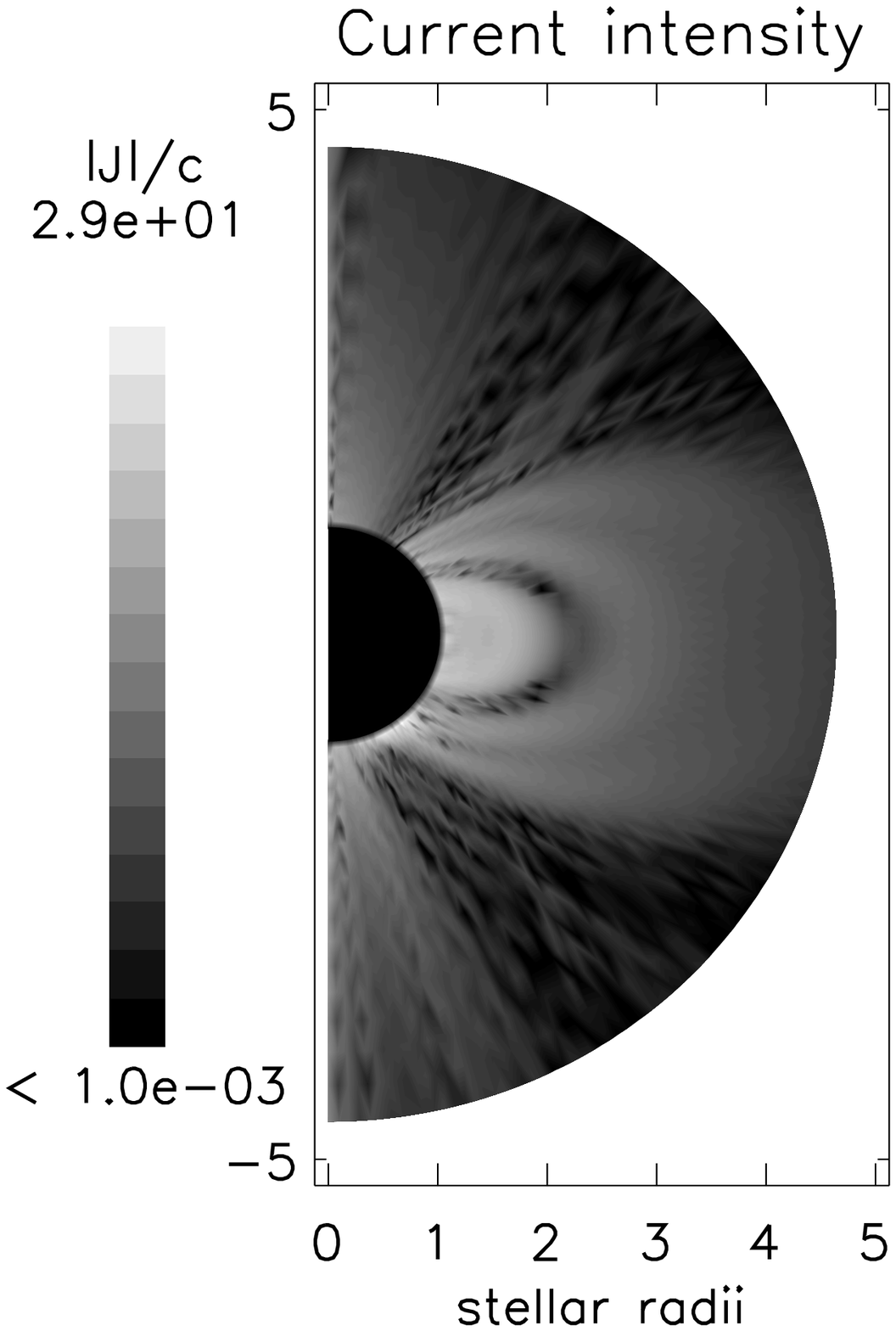}
\caption{Numerical model 1 (left) and model 2 (right). First panel: %$B_r(r_*)$, 
$|\vec B|$ (grey logarithmic scale) with scattering surfaces for photons of 1, 3 and $5~$keV (here $\beta=0.5$, $\mu=0.5$; see text); second panel: current intensity
$|\vec{J}|/c$ (grey linear scale), in units of $10^{14}~$G$/r_\star$.} \label{fig:model12}
\end{figure}

As an alternative to self-similar models, in \cite{vigano11} we presented a numerical  method to build general solutions. The force-free solutions are obtained by iteration starting from an arbitrary, non-force-free poloidal plus toroidal field. By the introduction of a fictitious velocity field proportional to the Lorentz force, the code is designed to dissipate only the part of the current not parallel to the MF. The boundary conditions are an arbitrary, fixed radial MF at the surface, $B_r(r_\star,\theta)$, and the requirement of a continuous matching with a potential dipole at large distances $r_{out}\ge 100 r_\star$. This method is able to produce general configurations, whose features may be radically different from the self-similar solutions. In this work, we try to reproduce some $j$-bundle like structure, aiming at providing input for MC simulations \cite{ntz08b}. In particular, we focus on two models, whose MF and current distribution are shown in Fig. \ref{fig:model12}. The first one is obtained fixing a dipolar form for radial MF, while the second one has a concentration of field lines near the south pole. To get these profiles, we have chosen the $\phi$ component of the vector potential as:
\begin{eqnarray}\label{aphi_models}
 && \mbox{model 1:} \qquad A_\phi(\theta,r_\star)=B_0\frac{\sin\theta}{2} \\
 && \mbox{model 2:} \qquad A_\phi(\theta,r_\star)=B_0\frac{\sin\theta}{2}\left[1+4e^{-\frac{(\theta-0.9\pi)}{0.2^2}}\right]
\end{eqnarray}
from which we recover $B_r=(\vec{\nabla}\times A_\phi\hat{\phi})_r$ ($B_0$ is the value of $B_r$ at north pole). Both models have high helicity, but it is not homogeneously distributed like in self-similar models. Model 1 is characterized by an equatorial asymmetry, with current and twist concentrated in a closed bundle near the equatorial region and a $j-$bundle near the southern semi-axis. Model 2 is more extreme: its helicity is $\sim$ 25 times larger than for model 1. Both are supposed to be more realistic than the self-similar models, since currents are more concentrated near the axis.

\section{Resonant Compton scattering}

The MC code we employ is presented in
\cite{ntz08a,ntz08b} and we refer to these works for further
details. Here we just summarize some basic concepts. It is assumed
that currents flow in a non-potential magnetosphere and that the
electron density is large enough for the medium to be optically
thick to resonant scattering: photons produced by the cooling star
surface can be then up-scattered by RCS, populating the hard tail
of the spectrum. In the stellar frame, the resonant energy
corresponds to the Doppler-shifted gyration frequency of the
electron
\begin{equation}
\omega_D(B,\beta)=\frac{1}{\gamma(1-\beta\mu)}\frac{eB}{m_ec}
\end{equation}
where $B=|\vec{B}|$ is the local intensity of MF, $c\beta$ the
electron velocity, $\gamma=1/\sqrt{1-\beta^2}$ the  Lorentz
factor, and $\mu$ the cosine of the angle between the photon
direction and the electron momentum. The non-resonant
contributions are negligible, since the resonant cross-section is
up to five orders of magnitude larger than the Thomson
cross-section. Figs. \ref{fig:model12} show the scattering
surfaces in the two models, for photons of 1, 3 and $5~$keV,
assuming $\beta=0.5$ and $\mu=0.5$. The surfaces are far from
being spherically symmetric. In model 2, which has the strongest
helicity, the scattering surfaces of soft X-ray photons lie tens
of stellar radii away from the surface, especially near the axis.

The scattering optical depth for a given seed photon energy depends on the local particle density $n_e$, the velocity of the scatterer and the local value of the MF. Since photons can be pushed to energies of some hundreds keVs, non-conservative scattering and the fully relativistic QED cross sections are used \cite{ntz08b}. The velocity distribution is assumed to be a one-dimensional (along the  field line), relativistic Maxwellian distribution with a plasma temperature $T_e$ and a bulk velocity $\beta_{bulk}$, as described in \cite{ntz08a}. The strong simplification in this approach is that the velocity distribution does not depend on position, which reduces the microphysical inputs to two parameters: the temperature of the plasma $T_e$ and $\beta_{bulk}$. Only recently, a more consistent treatment of the charges space and velocity distribution was attempted \cite{beloborodov11}. These new results show that the velocity distribution is non-trivial and strongly depends on position, which opens a new range of possibilities, so far unexplored.

Given a seed spectrum for the photons emitted from the surface
(assumed as a blackbody with  temperature $T_\star$), the MC code
follows the photon propagation through the scattering
magnetosphere. When a photon enters in a parameter region where no
more resonant scatterings are possible, the photon escapes to
infinity, where its energy and direction are stored. The sky at
infinity is divided into patches, so that viewing effects can be
accounted for (e.g. if line-of-sight is along the angles
$\theta_s$, $\phi_s$, only photons collected in the patch which
contains that pair of angles are analyzed). The angle-averaged
spectrum is obtained simply by averaging over all patches.

\section{Results and discussion}

\begin{figure}
\begin{minipage}[t]{3in}
\centering
\includegraphics[width=.95\textwidth]{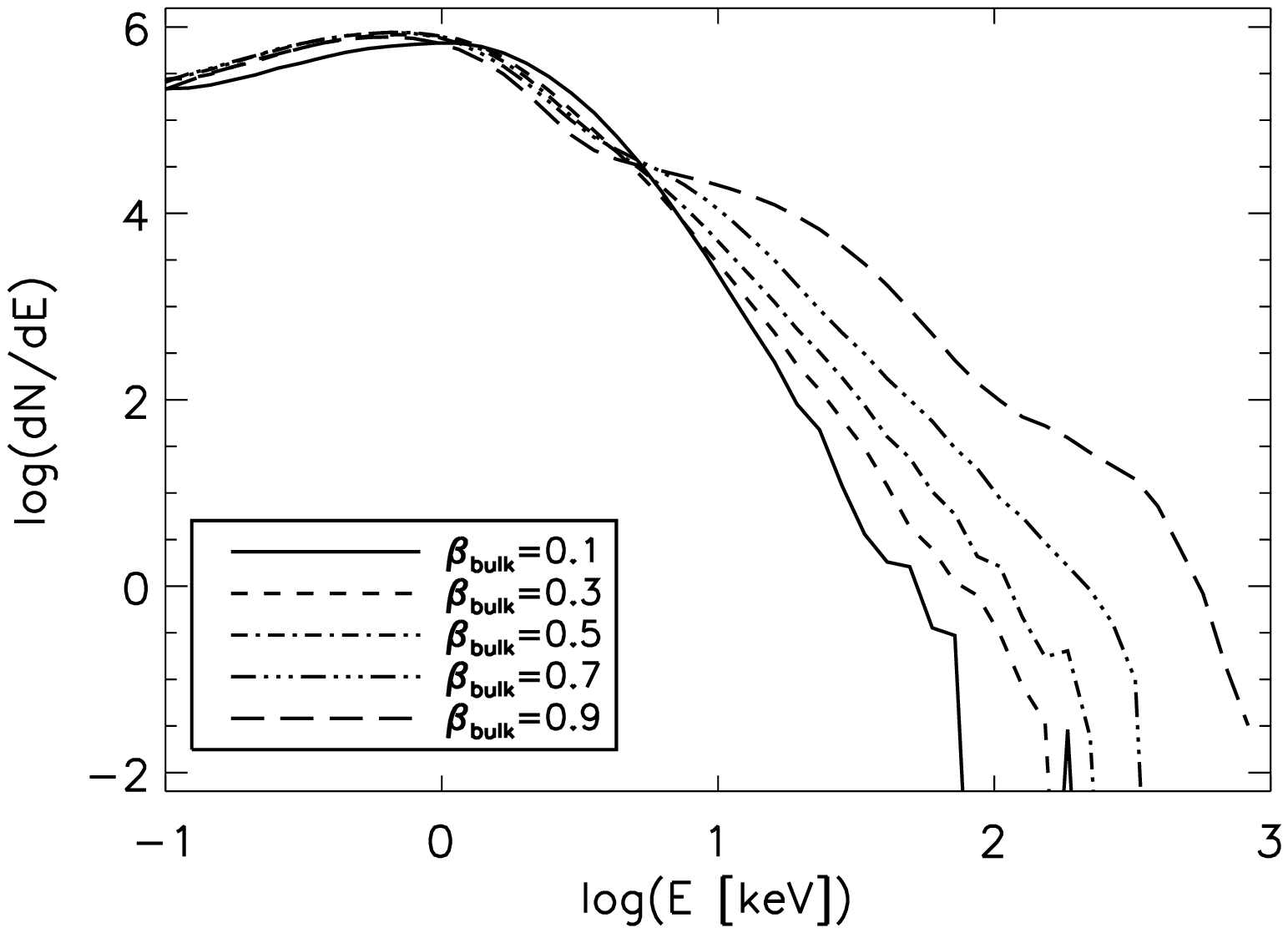}
\caption{Angle-averaged spectra for $kT_e=20~$keV, model 1, changing $\beta_{bulk}$.}
\label{fig:beta_bfinal1}
\end{minipage}
\hspace{0.5cm}
\begin{minipage}[t]{3in}
\centering
\includegraphics[width=.95\textwidth]{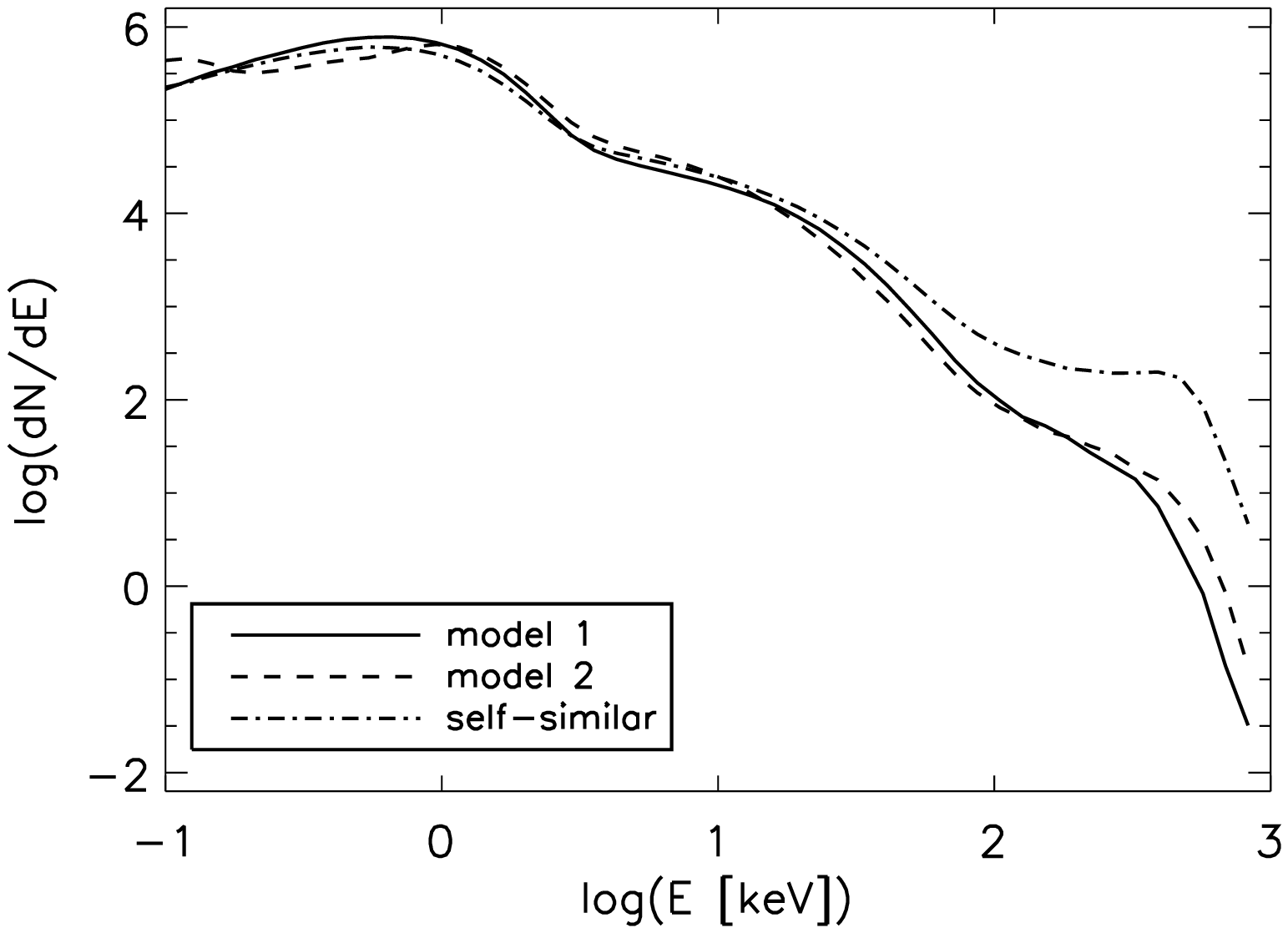}
\caption{Angle-averaged spectra for $kT_e=20~$keV, $\beta_{bulk}=0.9$, changing MF model.}
\label{fig:beta09}
\end{minipage}
\end{figure}

In previous works \cite{zane09}, self-similar magnetospheric
models were used to obtain synthetic spectra that were
satisfactorily fitted to X-ray observations, giving typical values
of $\beta_{bulk}\in[0.1,0.7]$  and $\Delta\phi\in[0.4,2.0]$. In
this paper we explore the dependence of the spectrum on the
magnetosphere model by comparing results obtained with
self-similar models and with our new numerical solutions for
force-free magnetospheres. We present results comparing models
with a magnetic field intensity at the pole of $B_p=10^{14}$~G.
The seed spectrum is assumed to be a $kT_\star=0.4~$keV Planckian
isotropic distribution for both ordinary and extraordinary photons. In Figs. \ref{fig:beta_bfinal1} and
\ref{fig:beta09} we plot the angle-averaged spectra, i.e., the
distribution of all photons escaped to infinity, independently on
the final direction. Fig. \ref{fig:beta_bfinal1} shows the effects
produced by changes in the bulk velocity, keeping fixed the MF
configuration (model 1), and the electron temperature ($kT_e=20~$keV). The main fact is the the spectrum in the region $E>10~$keV
becomes harder as  $\beta_{bulk}$ is increased, while it is
depleted of photons of energies in the $1-10~$keV range. A similar,
but less pronounced effect, is obtained by increasing the electron
temperature \cite{ntz08b}.

In Fig. \ref{fig:beta09} we compare spectra obtained with fixed
values of $kT_e=20~$keV and $\beta_{bulk}=0.9$, varying only the
MF topology. The three lines correspond to  model 1 (solid), model
2 (dashes) and a self-similar model (dash-dotted line) with
$\Delta\phi=1.36$, which approximately has the same total helicity
of model 1. We have chosen a high value of $\beta_{bulk}$ to show
a case where the effects are larger, but our conclusions do not
change qualitatively for lower values of $\beta_{bulk}$. The
high-energy cutoff produced by electron recoil is also evident. We
see that the major differences arising from changes in MF topology
appear in the hard tail of the spectra, which is clearly harder
for the self-similar model. However, the thermal part of the spectrum is more depleted,
especially for model 2.

\begin{figure}
\centering
\includegraphics[width=.47\textwidth]{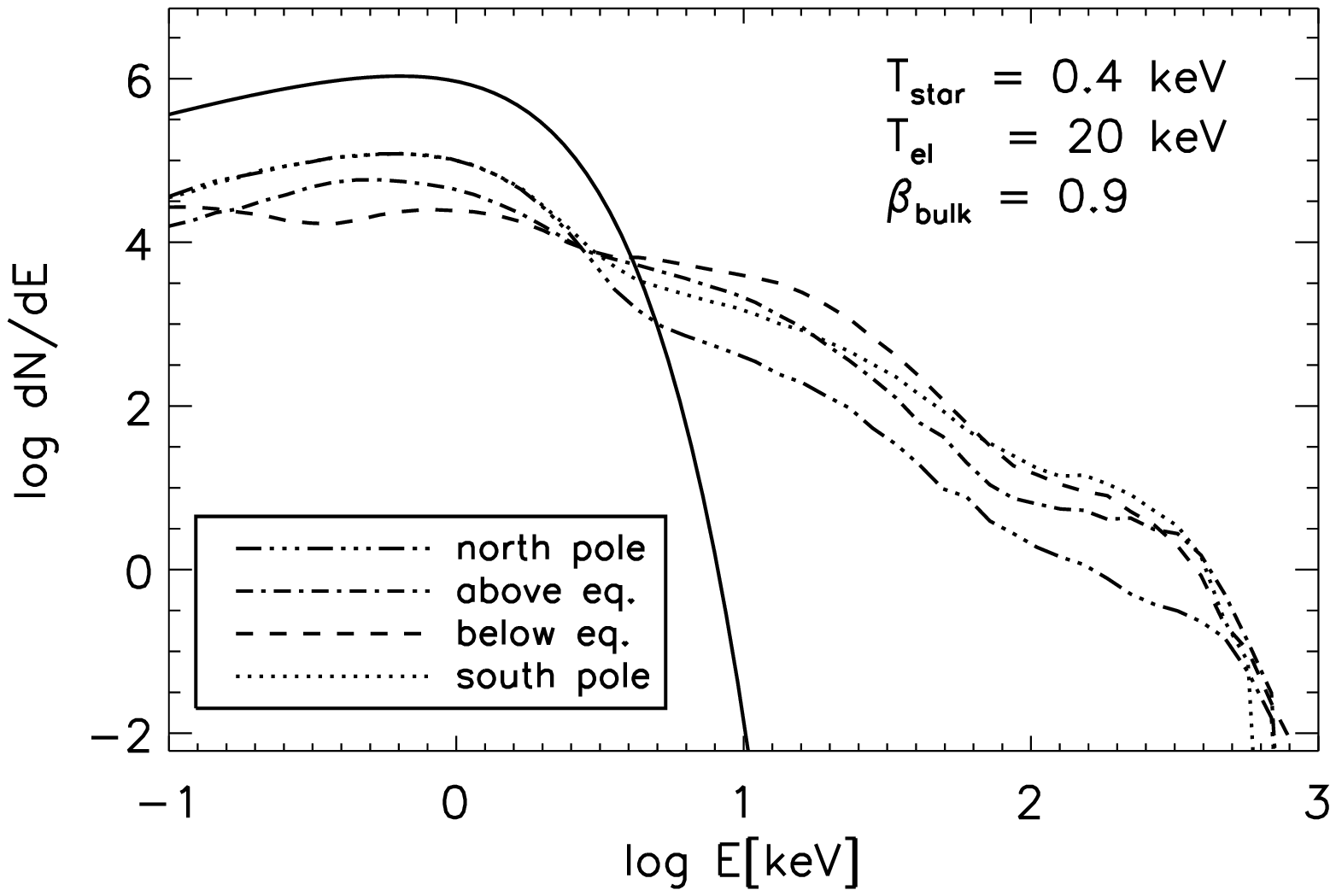}
\includegraphics[width=.47\textwidth]{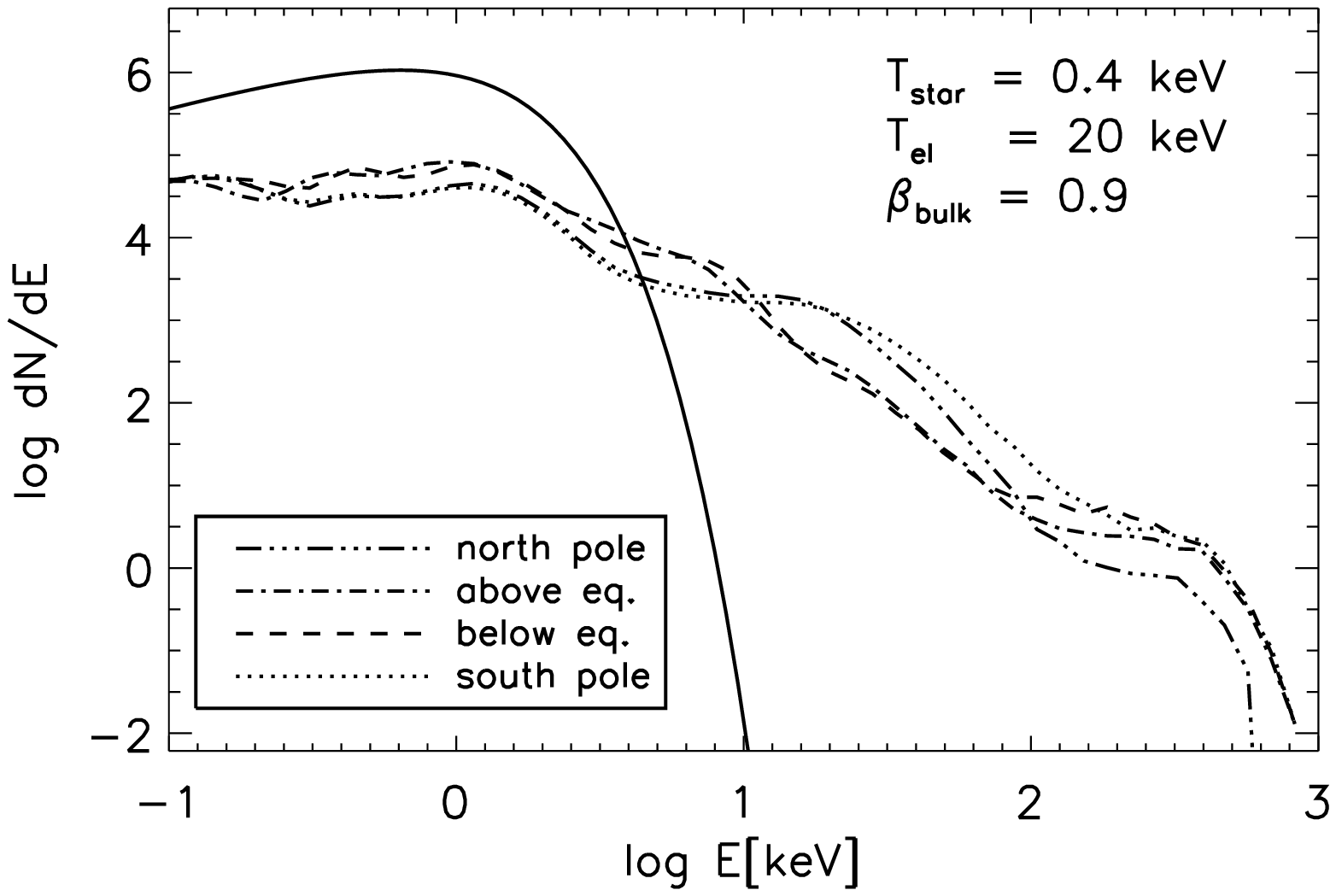}\\
\medskip
\includegraphics[width=.47\textwidth]{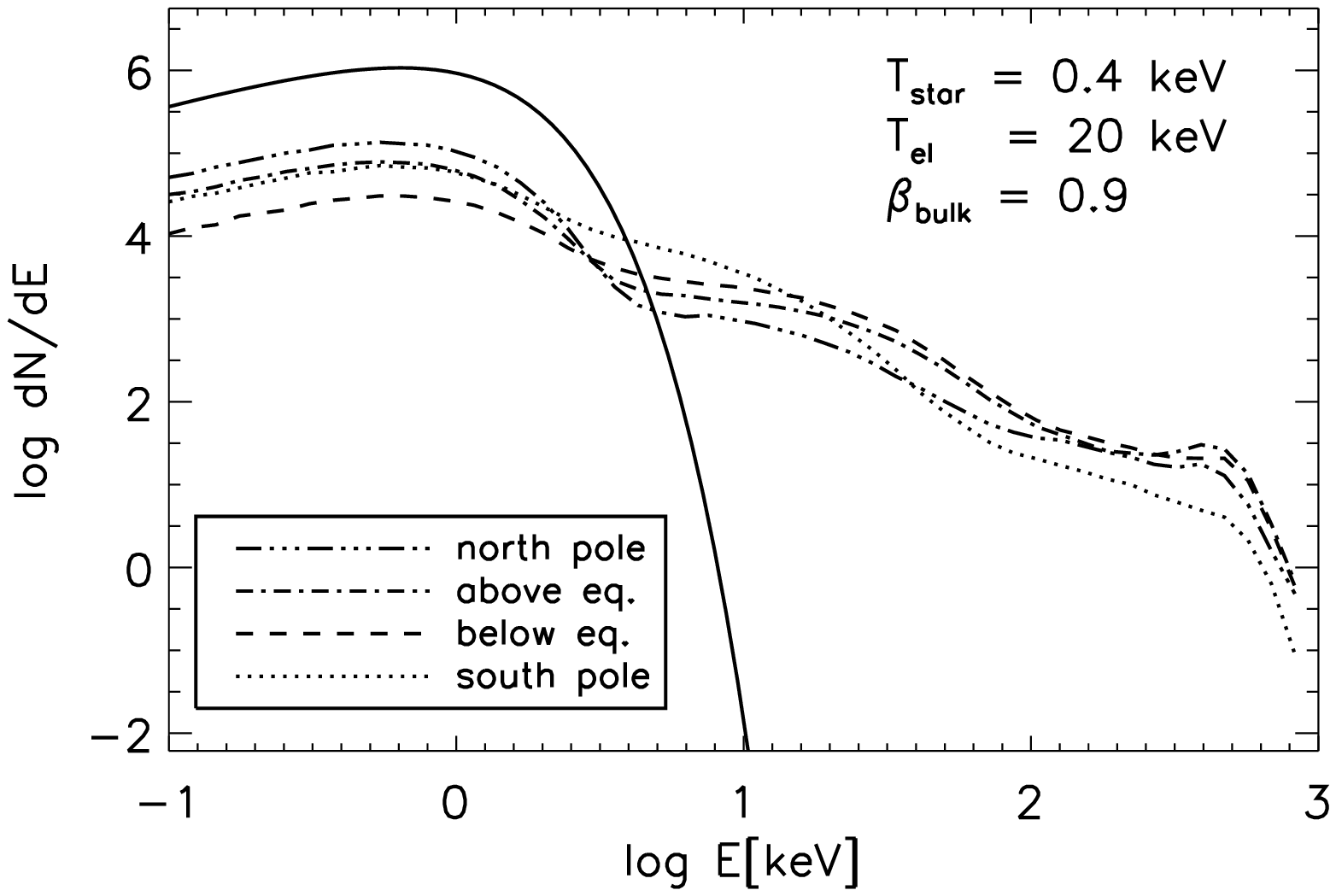}
\includegraphics[width=.47\textwidth]{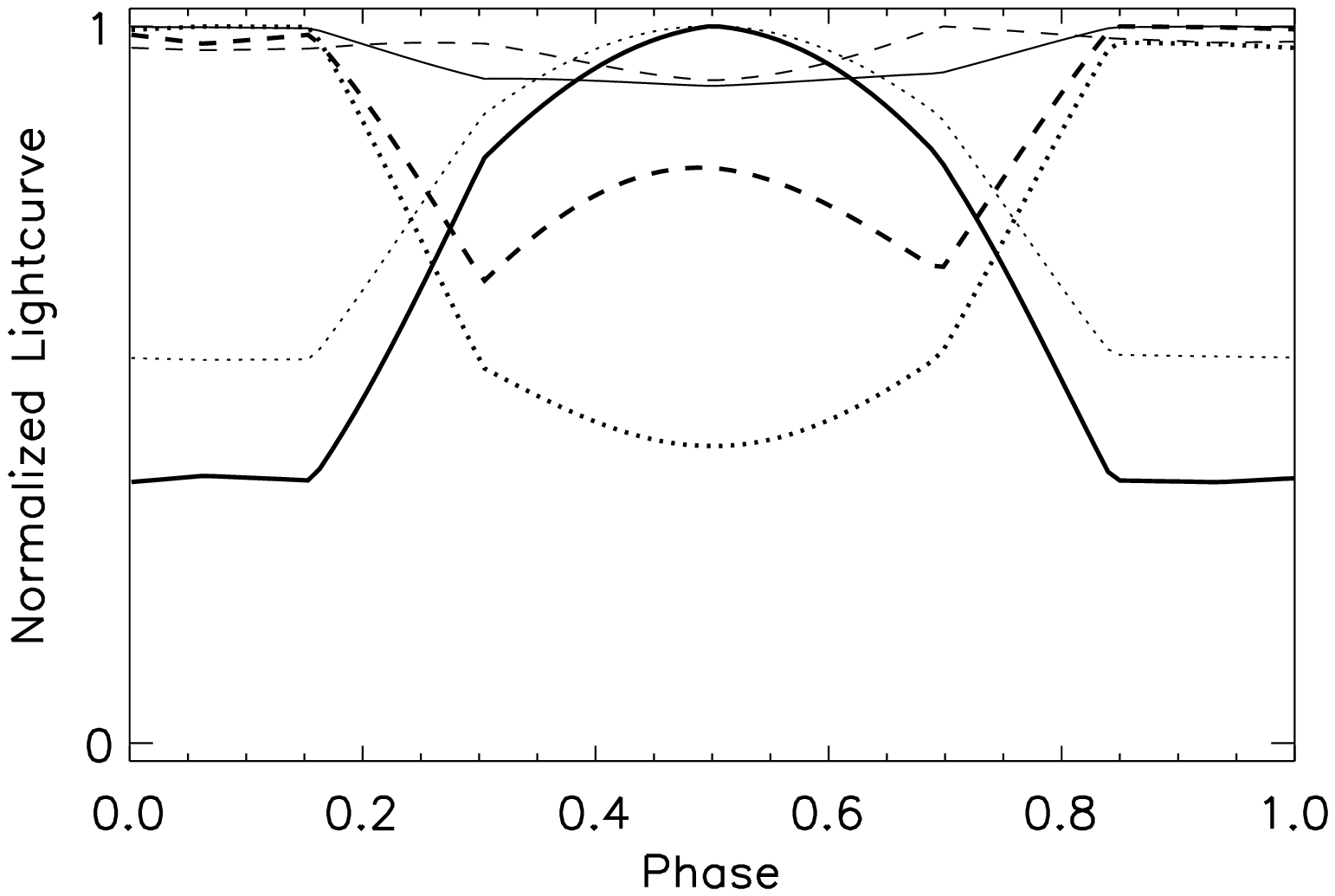}
\caption{Synthetic spectra for model 1 (upper left panel), model 2 (upper right), self-similar model (lower left) with the parameters indicated in the figures. Legends indicate the four viewing angles here considered: aligned with north or south pole, just above and just below the equator. The seed blackbody is shown for comparison (solid line). Fourth panel shows the pulse profiles obtained for model 1 (solid line), model 2 (dotted), self-similar (dashed), in the energy ranges 0.5--10 keV (thin lines) and 20--200 keV (thick lines), with $\xi=\chi=30^\circ$ (see text).}
\label{fig:dir}
\end{figure}

However, in the angle-averaged spectrum, the differences due to
the MF topology are partially smeared out, and the comparison
between synthetic spectra as seen from different viewing angles
(first three panles in Fig. \ref{fig:dir}) is more interesting. The comptonization
degree reflects the inhomogeneous particle density distribution
and therefore  the particular current distribution produces
important differences between different viewing angles. Since
neutron stars rotate, the study of pulse profiles and
phase-resolved spectra can trace the geometric features of the
scattering region (tens of stellar radii). In the lower right panel we show the light curves for the three MF models, in bands 0.5--10~keV and 20--200~keV, for an oblique rotator with a certain line of sight ($\xi=\chi=30^\circ$ in \cite{ntz08a}).

For self-similar models (lower left panel), the thermal part of the spectrum shows a smooth dependence with the viewing angle, which in turn translates into relatively regular light curves. In models 1 and 2, differences in the spectra induced by viewing angles are more pronounced. In these cases, the spectrum is much
more irregular and asymmetries are larger. In particular, model 1 (upper left) shows a softer spectrum when seen from northern latitudes, with important
spectral differences. The pulsed fraction of model 1 is very high in the hard X-ray band, while it is comparable with the self-similar model in the soft range. Model 2 (upper right) has a more symmetric distribution of currents, and the comptonization degree at different angles depends on the energy band in a non trivial way. This results in large pulsed fraction for both energy ranges and in notable differences between their pulse profiles.

The high energy tail, seen at different colatitudes, can vary by one order of magnitude or more. Investigating the geometry can help to recognize the different components seen in the hard tails via pulse phase spectroscopy. On the other hand, the macrophysical approach used in this work to obtain the MF topology should be accompanied by the corresponding microphysical description to determine the velocity distribution of particles, here simply taken as a model parameter. A fully consistent description of both MF geometry and the spatial and velocity distribution of particles is needed to advance our interpretation of magnetar spectra.


\section*{References}
\begin{thebibliography}{9}
\bibitem{beloborodov07} Beloborodov A M and Thompson C 2007 {\it ApJ} {\bf 657} 967
\bibitem{beloborodov09} Beloborodov A M 2009 {\it ApJ} {\bf 703} 1044
\bibitem{beloborodov11} Beloborodov A M 2011 {\it High-Energy Emission from Pulsars and their Systems} ed Torres D F and Rea N p 299
\bibitem{fernandez07} Fern{\'a}ndez R and Thompson C 2007 {\it ApJ} {\bf 660} 615
\bibitem{fernandez11} Fern{\'a}ndez R and Davis S W 2011 {\it ApJ} {\bf 730} 131
\bibitem{low90} Low B C and Lou Y Q 1990 {\it ApJ} {\bf 352} 343
\bibitem{mereghetti08} Mereghetti S 2008 {\it A\&ARv} {\bf 15} 225
\bibitem{ntz08a} Nobili L, Turolla R and Zane S 2008 {\it MNRAS} {\bf 386} 1527
\bibitem{ntz08b} Nobili L, Turolla R and Zane S 2008 {\it MNRAS} {\bf 389} 989
\bibitem{pavan09} Pavan L, Turolla R, Zane S and Nobili L 2009 {\it MNRAS} {\bf 395} 753
\bibitem{pons11} Pons J A and Perna R 2011 {\it ApJ} {\bf 741} 123
\bibitem{tlk02} Thompson C, Lyutikov M and Kulkarni S R 2002 {\it ApJ} {\bf 574} 332
\bibitem{turolla11} Turolla R, Zane S, Pons J A, Esposito P and Rea N 2011 {\it ApJ} {\bf 740} 105
\bibitem{vigano11} Vigan\`o D, Pons J and Miralles J A 2011 {\it A\&A} {\bf 533} A125
\bibitem{zane09} Zane S, Rea N, Turolla R and Nobili L 2009 {\it MNRAS} {\bf 398} 1403

\end{thebibliography}
\end{document}